\documentclass[twocolumn,tighten]{aastex63}

\usepackage{graphicx}
\usepackage{amsmath}
\usepackage{nccmath}
\usepackage{comment}
\usepackage{tabularx}
\usepackage{algorithm2e}
\usepackage{float}
\usepackage{fullwidth}
\usepackage{nccmath}
\usepackage{mathtools}
\usepackage{multirow}
\usepackage{scrextend}
\usepackage{lineno}
\usepackage{float} 
\definecolor{yscol}{rgb}{0.8, 0.6, 1}

\newcommand{\msun}{\,{\rm M}_\odot}

\newcommand{\s}{\,{\rm s}}
\newcommand{\g}{\,{\rm g}}

\newcommand{\om}{\Omega_{\rm m}}
\newcommand{\sig}{\sigma_{8}}
\newcommand{\asnone}{A_{\rm SN1}}
\newcommand{\asntwo}{A_{\rm SN2}}
\newcommand{\aagnone}{A_{\rm AGN1}}
\newcommand{\aagntwo}{A_{\rm AGN2}}
\newcommand{\cov}{{\rm cov}}
\newcommand{\var}{{\rm var}}
\newcommand{\bSigma}{\boldsymbol{\Sigma}}
\newcommand{\bmu}{\boldsymbol{\mu}}
\newcommand{\btheta}{\boldsymbol{\theta}}
\newcommand{\bX}{\boldsymbol{X}}
\newcommand{\pd}{{\partial}}

\newcommand{\bx}{\boldsymbol{x}}

\newcommand{\dd}{\mathrm{d}}

\definecolor{yscol}{rgb}{0.8, 0.6, 1}
\def\ys#1{{\textcolor{yscol} {[#1]}}}
\def\bt#1{{{\bf #1}}}
\def\bs#1{{\boldsymbol{#1}}}

\newlength{\Oldarrayrulewidth}

\makeatletter
\newcommand{\thickhline}{%
    \noalign {\ifnum 0=`}\fi \hrule height 1pt
    \futurelet \reserved@a \@xhline
}
\newcommand{\thichline}{%
    \noalign {\ifnum 0=`}\fi \hrule height 0.8pt
    \futurelet \reserved@a \@xhline
}
\newcolumntype{"}{@{\hskip\tabcolsep\vrule width 0.8pt\hskip\tabcolsep}} 

\makeatother

\received{June 1, 2019}
\revised{January 10, 2019}
\accepted{\today}
\submitjournal{ApJ}

\graphicspath{{./}{figures/}}

\begin{document}


\title{On the Significance of Covariance for Constraining Theoretical Models From Galaxy Observables}

\correspondingauthor{Yongseok Jo}
\email{yjo@flatironinstitute.org}
\author[0000-0003-3977-1761]{Yongseok Jo}
\affiliation{Columbia Astrophysics Laboratory, Columbia University, New York, NY 10027, USA}
\affiliation{Center for Computational Astrophysics, Flatiron Institute, 162 5th Avenue, New York, NY, 10010, USA}

\author{Shy Genel}
\affiliation{Center for Computational Astrophysics, Flatiron Institute, 162 5th Avenue, New York, NY, 10010, USA}
\affiliation{Columbia Astrophysics Laboratory, Columbia University, 550 West 120th Street, New York, NY, 10027, USA}

\author{Joel Leja}
\affiliation{Department of Astronomy \& Astrophysics, The Pennsylvania State University, University Park, PA 16802, USA}
\affiliation{Institute for Computational \& Data Sciences, The Pennsylvania State University, University Park, PA 16802, USA}
\affiliation{Institute for Gravitation and the Cosmos, The Pennsylvania State University, University Park, PA 16802, USA}

\author{Benjamin Wandelt}
\affiliation{Sorbonne Universite, CNRS, UMR 7095, Institut d’Astrophysique de Paris, 98 bis boulevard Arago, 75014 Paris, France}
\affiliation{Center for Computational Astrophysics, Flatiron Institute, 162 5th Avenue, New York, NY, 10010, USA}





\begin{abstract}
In this study, we investigate the impact of covariance within uncertainties on the inference of cosmological and astrophysical parameters, specifically focusing on galaxy stellar mass functions derived from the CAMELS simulation suite.
Utilizing both Fisher analysis and Implicit Likelihood Inference (ILI), we explore how different covariance structures, including simple toy models and physics-motivated uncertainties, affect posterior distributions and parameter variances.
Our methodology utilizes forward modeling via emulators that are trained on CAMELS simulations to produce stellar mass functions based on input parameters, subsequently incorporating Gaussian noise as defined by covariance matrices. 
We examine both toy model covariance matrices and physically motivated covariance matrices derived from observational factors like the stellar Initial Mass Function (IMF) and photometric aperture size. 
Our results demonstrate that covariance terms significantly influence parameter inference, often leading to tighter constraints or revealing complex, multimodal posterior distributions. 
These findings underscore the necessity of accounting for covariance when interpreting astrophysical observations, especially in fields where accurate parameter estimation is critical for model validation and hypothesis testing.
\end{abstract}


\section{Introduction}
The cornerstone of scientific progress lies in the ability to make precise measurements and draw reliable conclusions from observed phenomena. 
However, no measurement or observation in science is complete without a thorough understanding and quantification of its associated uncertainties. 
Precise quantification of uncertainty and confidence levels is of paramount importance in practicing science.
This is crucial not only for validating experimental results but also for forming the foundation of every theoretical model.

In an era of extensive and precise observation, the challenge of accurately quantifying uncertainty has become both more crucial and more complex.
This is particularly true in fields such as cosmology and astrophysics, where observations are often indirect, subject to various systematic effects, and limited by the fundamental nature of studying distant phenomena.
The confidence with which we can make claims about the universe's structure, composition, and evolution hinges on our ability to rigorously account for uncertainties in our measurements and models.
Moreover, confidence levels play a critical role in hypothesis testing and model validation.



The calibration of galaxy formation models is one of the significant and challenging examples of model validation.
Over the decades, considerable efforts have been made to replicate the observable universe using computational methods, such as cosmological (magneto-) hydrodynamic simulations or semi-analytic models, in conjunction with dark matter-only simulations \citep{illustris2014MNRAS.444.1518V,genel2014MNRAS.445..175G,schaye2015MNRAS.446..521S,magneticum2017A&C....20...52R,nelson2018MNRAS.475..624N,dave2019MNRAS.486.2827D,universemachine2019MNRAS.488.3143B,santacruzsam2022MNRAS.517.6091G,astrid2022MNRAS.513..670N}. 
Calibration of these models has traditionally been done manually, often involving basic visual comparison with observations, though some relatively fast models have also been calibrated against observations using Monte Carlo Markov Chain methods \citep{universemachine2019MNRAS.488.3143B}. 
More recently, there has been a growing effort to adopt more mathematically rigorous calibration techniques that achieve computational feasibility by leveraging machine learning methodologies \citep{Jo2023ApJ...944...67J, flamingo2023MNRAS.526.6103K}.
Nevertheless, these methodologies have encountered challenges in accurately accounting for uncertainties, particularly neglecting the covariance terms of the uncertainty, specifically the off-diagonal elements of the uncertainty matrix.

In contrast, the importance of covariance matrix has been explored in various contexts.
Particularly in the field of cosmology, numerous studies have been conducted concerning the covariance matrix.
For instance, \citet{ereza2024MNRAS.532.1659E} investigates the covariance of two-point correlation functions through cosmological simulations.
\citet{stone2021ApJ...912...41S} aims to refine the intrinsic scatter of scaling relations by analyzing covariance between different observables.
\citet{yang2023MNRAS.519.4938Y} propose that exploiting covariance between data points using Gaussian processes can provide improved constraints on the Hubble constant, while \citet{posti2022RNAAS...6..233P} employs Gaussian processes to infer mass and concentration from rotation curves. 
On the other hand, \citet{marchesini2009ApJ...701.1765M} provides a comprehensive analysis of correlated observational uncertainties concerning the galaxy stellar mass function, and \citet{pacifici2023ApJ...944..141P} addresses observational discrepancies between different fitting models and their consensus, hinting at the potential need for rigorous uncertainty construction that accounts for these discrepancies.
However, none of these studies specifically address the impact of physics-driven covariance within uncertainty of observables related to the galaxy populations on inference or parameter estimation.

This work focuses on estimating the effect of uncertainty covariance of the galaxy stellar mass functions, which has been neglected mainly due to practical constraints like computational cost. 
The covariance associated with the galaxy stellar mass functions corresponds to the correlation between distinct pairs of observed data points, specifically different stellar mass bins.
The covariance typically arises when analyzing through physical models the observational datasets to extract physical quantities, such as galaxy stellar masses and galaxy star formation rates. 
For instance, the covariance can be constructed by running multiple iterations of Spectral Energy Distribution (SED) fitting with varying physical parameters such as some parameters of initial mass functions (IMF), which is in general computationally heavy.
This issue is circumvented by considering the effect of IMFs on the galaxy stellar mass function to be generic (refer to Sec. \ref{sec:method_observation} for details).

Another challenge is the computational cost associated with theoretical models that produce the galaxy stellar mass functions, such as cosmological hydrodynamic simulations. 
To tackle this, we employ an emulator trained on the CAMELS simulations as a forward model for generating galaxy stellar mass functions from input parameters \citep{Jo2023ApJ...944...67J}.
With this, we utilize a two-pronged approach to inference.
Firstly, we apply Fisher analysis, a technique to compute the information content of data by bounding the covariance matrix of optimal parameter estimators.
This involves constructing a Fisher information matrix and estimating variance using the Cramér-Rao bound. Secondly, we implement Implicit Likelihood Inference (ILI), a machine learning-based method that trains a neural network to estimate posterior distributions within a Bayesian framework. By employing both techniques, we aim to cross-validate our findings and compare the outcomes of these distinct inferential approaches.

To incorporate uncertainty correlations in galaxy stellar mass functions, we formulate covariance matrices based on the assumption of a multivariate Gaussian distribution. 
This formulation results in a multivariate Gaussian function characterized by a mean galaxy stellar mass function with a covariance matrix, subsequently generating stellar mass functions with corresponding uncertainties.
Our approach begins with three conceptually simple covariance matrices: fully correlated, partially correlated, and uncorrelated. This allows us to examine how varying degrees of correlation in uncertainty influence the inference process.
We then progress to more sophisticated, physically motivated covariance matrices derived from actual observational data. These matrices integrate correlation information about IMF parameters in Spectral Energy Distribution (SED) fitting, photometric aperture size, and sampling. This enables us to investigate the impact of complex, real-world covariances on the results.
Through this comprehensive methodology, we aim to provide a conceptual understanding of how correlations within uncertainty affect parameter estimation and model calibration in astrophysical contexts.

The structure of this paper is as follows. 
In Sec. \ref{sec:method}, we provide the overview of Fisher analysis (Sec. \ref{sec:method_fisher}) and implicit likelihood inference (Sec. \ref{sec:method_ili}) that are used to conduct inference.
In Sec. \ref{sec:method_sim_param} and \ref{sec:method_observation}, we describe the model and parameters used for inference and observation used to construct covariance matrices.
In Sec. \ref{sec:toy_models}, we investigate the impact of covariance matrices on inference using simple toy models. 
In Sec. \ref{sec:phys_model}, we construct physically-motivated covariance matrices from the actual observations and investigate the impact of the the covariance matrices on the inference.   
Lastly, we present a summary of the results and findings in Sec. \ref{sec:sum}.\\

\section{Methodology}
\label{sec:method}
\subsection{Fisher Analysis}
\label{sec:method_fisher}
The Fisher information is a mathematical tool for estimating the amount of information that a random variable $X$ carries about parameters $\theta$.
The random variable $X$ is a function that maps the outcomes of a random process $\Omega$ to a numeric value space $E$ (i.e., ${X}:\Omega\rightarrow E$).
For the simplest example, considering coin flipping, the possible outcomes are the head and tail (i.e., $\Omega={\mathrm{Head}, \mathrm{Tail}}$, while it can be expressed as 0 (head) or 1 (tail) (i.e., $E={0,1}$).
Then, $P(X=0)$ represent the probability of a head.
In this work, the random process is cosmological simulations with cosmological parameters and sub-grid parameters $\btheta$.
The possible outcomes are various galaxy populations across the simulated universe.
Subsequently, the galaxy stellar mass functions, as an array of counts per mass bin, can be represented by $\bs{X}$, typically conditioned on $\btheta$.

The Fisher matrix can be written as 
\begin{equation}
\label{eq:Fisher}
\mathcal{F}_{ij} = \mathbb{E}\left[\left(\frac{\partial}{\partial\theta_i}\log\left(f(\bX;\btheta)\right)\right)\left(\frac{\partial}{\partial\theta_j}\log(f(\bX;\btheta))\right)\big|\btheta\right],
\end{equation}
where $f(\bs{X}; \bs{\theta})$ is the probability distribution of $\bs{X}$ conditioned on $\btheta$.
Under the assumption that the probability distribution is the multivariate
 Gaussian ($\bX\sim \mathcal{N}(\bmu(\btheta),\bSigma)$) and the covariance matrix $\bSigma$ is independent of the parameters ($\partial \bs{\Sigma} / \partial \theta_i = 0$), Eq.~\ref{eq:Fisher} simplifies to
\begin{equation}
\label{eq:Fisher_simplified}
    \mathcal{F}_{ij} = \frac{\partial\bmu^{\rm T}}{\partial\theta_i}\bSigma^{-1}\frac{\partial\bmu}{\partial\theta_j}. 
\end{equation}
With this, the Cramer-Rao bound provides a lower bound on the variance of unbiased estimators\footnote{An unbiased estimator averages to the true value of the parameter in the limit when the experiment is repeated an infinite number of times. 
In other words, the expected value of an unbiased estimator is equal to the true value of the parameter.} of a parameter with the following inequality:
\begin{equation}
\label{eq:cramer_rao}
\var(\theta_{ii}) \geq (\mathcal{F}^{-1})_{ii}.
\end{equation}
Here, $\var(\theta_{ii})$ is a marginal variance, not conditional. 
The conditional variance can be expressed as $(F_{ii})^{-1}$, by using only one components of the Fisher matrix while excluding the other components of parameters.
This is analogous to fixing the parameters (i.e., $d\theta_j=0$) during the computation of the Fisher matrix in Eq. \ref{eq:Fisher_simplified}. 
Conversely, employing the Fisher matrix as whole results in the marginal variance, accounting for uncertainty associated with all the parameters.


\subsection{Implicit Likelihood Inference}
\label{sec:method_ili}
Implicit likelihood inference (ILI) is an inference method that requires no explicit formalism for likelihood functions. 
In general, this can be achieved by employing a neural network that can estimate a probability function as likelihood \citep{nde2016arXiv161003483M,papamakarios2018arXiv180507226P,Alsing_2018,Cranmer2020,durkan2020arXiv200203712D}. 
This facilitates, e.g., Bayesian inference without writing down the explicit likelihood in any analytic form.

The fundamental concept of ILI is described as follows.
Given the true (ideal) posterior distribution $p(\bx|\btheta)$ and the empirical posterior distribution $\tilde{p}(\bx|\btheta)$ that can be derived from the dataset pairs $(\bx,\btheta)$, as the dataset size approaches infinity, $\tilde{p}(\bx|\btheta)\rightarrow p(\bx|\btheta)$. 
To infer $\tilde{p}(\bx|\btheta)$ from the data, we employ a neural density estimator (NDE) that takes $\bx$ and $\btheta$ as input and outputs the probability $q_\phi(\bx,\theta)$. 
Subsequently, we train the NDE utilizing the following loss function:
\begin{equation}
\begin{split}
    \mathcal{L} &= -\mathbb{E}_{\tilde{p}(\bx,\btheta)}\big[\log(q_{\phi}(\bx,\btheta))\big]\\ &\sim \mathbb{E}_{p(\btheta)}\big[\mathrm{KL}\big(\tilde{p}(\bx|\btheta)\big|\big|q_{\phi}(\bx,\btheta)\big)\big] + \mathrm{constant},
\end{split}
\end{equation}
where $\phi$ stands for hyperparameters to be optimized during training, and $\mathrm{KL}(P||Q)$ denotes the Kullback–Leibler divergence, a measure of similarity (or discrepancy) between two probability distributions $P$ and $Q$, giving non-negative real and equal to 0 if and if only $P=Q$. 
In this context, $\mathbb{E}_{p}[q]$ calculates the expectation value of $q$ based on the dataset, which is sampled according to the probability distribution $p$.
This quantifies the cross entropy (i.e., $H = \sum p\cdot \log(q)$), providing an estimate of the similarity between the two distributions.
Therefore, by minimizing the loss function---minimizing the distance between $\tilde{p}$ and $q_\phi$, the NDE $q_\phi$ approximates the the empirical posterior distribution $\tilde{p}(\bx|\btheta)$ given the prior $p(\theta)$.
Notably, the precision attained is contingent upon the training and validation datasets.
In addition, as the size of data increases, $\tilde{p}$ asymptotically approaches the true $p$.

\subsection{Forward Model}
\label{sec:method_sim_param}
In this work, we focus on inference of cosmological and astrophysical parameters from a stellar mass function.
To this end, we use an emulator trained on the cosmological simulations of the CAMELS project to estimate stellar mass, taking cosmological and astrophysical parameters as input.

The CAMELS project explored a wide range of cosmological and astrophysical parameters through thousands of cosmological simulations with a comoving volume of $(25 h^{-1}\mathrm{Mpc})^3$ \citep{camels2021ApJ...915...71V}.
Amongst the suites of the CAMELS simulation, we use the LH set run with the IllustrisTNG engine.
The LH set consists of 1000 simulations with variations of two cosmological parameters ($\om$, $\sig$), two stellar feedback parameters ($\asnone$, $\asntwo$), and two AGN feedback parameters ($\aagnone$, $\aagntwo$) (refer to Sec. 3.1 in \citet{camels2021ApJ...915...71V} for more details).
The ranges of parameter space of the LH set are as follows: $\Omega_\mathrm{m} \in [0.1, 0.5]$, $\sigma_8 \in [0.6,1.0]$, $A_\mathrm{SN1}\in [0.25, 4.0]$, $A_\mathrm{SN2}\in [0.5,2.0]$, $A_\mathrm{AGN1}\in [0.25,4.0]$, and $A_\mathrm{AGN2}\in [0.5,2.0]$.

Although the CAMELS simulations cover a wide range of parameter space, it is insufficient to conduct a full inference with only existing simulations. 
For this reason, we adopt the emulator from \citet{Jo2023ApJ...944...67J} that outputs stellar mass functions with 13 mass bins ranging from $10^{8.9}$ to $10^{11.4}\msun$.
The emulator is assumed to be  unbiased and to output the expected value $\bs{\mu}$ in Eq. \ref{eq:Fisher_simplified}. 
The stellar mass function with uncertainty, denoted as $\bs{x}$, can be obtained by using the Gaussian noise as $\bX\sim \mathcal{N}(\bmu(\btheta),\bSigma)$ given covariance matrix $\bSigma$.
With the sampled ($\bs{x}$, $\bs{\theta}$), we can train the NDE. 
In the case of the Fisher analysis, $\partial\bs{\mu}/\partial\theta_i$ is simply calculated numerically using the finite difference method.

The target stellar mass function, against which inference is conducted throughout this paper, is generated by the emulator with the fiducial parameters of the CAMELS simulations as input, which is  ($\om$, $\sig$, $\asnone$, $\asntwo$, $\aagnone$, $\aagntwo$) = (0.3,0.8,1,1,1,1).
For inference, uniform priors are employed consistent with the parameter ranges of the CAMELS project.

\subsection{Observational Data}
\label{sec:method_observation}
Here we perform stellar population synthesis modeling in order to assess the effect of an unknown IMF on the inferred mass for a single galaxy. This posterior will later be used to simulate covariant uncertainties in the inferred stellar mass function. This galaxy is GOODS-S 40305, a typical star-forming galaxy at $z=1.61$. The data comes from the 3D-HST photometric catalog, which provides observed-frame photometry covering the UV to the mid-IR for tens of thousands of distant galaxies \citep{skelton2014ApJS..214...24S}. The galaxy is fit using Prospector \citep{johnson2021ApJS..254...22J}, using the Prospector-$\alpha$ model \citep{leja2019ApJ...877..140L}.

Only a single object is fit because the building blocks of these fits, simple stellar populations, require an IMF as input \citep[e.g.,][]{conroy2013ARA&A..51..393C}. Constructing new simple stellar populations from the underlying stellar libraries and isochrones takes several minutes on a modern laptop. As fitting one galaxy typically requires millions of model evaluations, it would be computationally expensive to recompute the simple stellar populations on-the-fly while sampling. We therefore instead combine posteriors from multiple independent fits with fixed (different) IMFs to create a single effective posterior. As this is a computationally expensive procedure, at tens of hours per fit \citep{leja2019ApJ...877..140L}, we accordingly only do it for one object. Nonetheless this is a representative result as changes to the IMF will typically have a similar multiplicative effect on the total stellar mass of a galaxy, regardless of its SED.

In this exercise we take a standard Kroupa IMF \citep{kroupa2001MNRAS.322..231K}, and assume that the slope for stars with mass $M^* > 0.5$ M$_{\odot}$ is constant and can vary between 1.3 and 3.3 (a uniform prior), compared to the standard Salpeter slope of $\alpha = 2.35$ \citep{salpeter1955ApJ...121..161S}. We also allow the high-mass cutoff to vary between 100 and 300 M$_{\odot}$ (compared to the default value of 100 M$_{\odot}$. We then run 100 separate fits in a two-dimensional grid of these values, each with a fixed IMF, using the nested sampling algorithm \texttt{dynesty} \citep{speagle2020MNRAS.493.3132S}. After the fit, we combine the posteriors using Bayesian model averaging, which effectively weights each posterior sample by the Bayesian evidence. In this way we can effectively sample over the IMF without changing the simple stellar populations during the fit.

The covariance between total stellar mass and the power-law slope and high-mass cutoff of the IMF is later assumed to be universal and used to construct an estimate of the covariance between bins of the stellar mass function.

\section{Posterior Sensitivity to the Covariace: Analytical Toy Models}
\label{sec:toy_models}
We investigate how different uncertainties, characterized by various covariance structures between the bins of the mass function, affect the inferred posterior distributions of the simulation parameters. 
To ensure robustness of results and compare methodologies, we perform inference using both Fisher analysis and implicit likelihood inference (ILI). 
This dual approach allows us to validate our results across different methods while simultaneously highlighting any discrepancies between them.

We employ three simple covariances that can be written analytically: (a) fully uncorrelated uncertainty, (b) partially correlated uncertainty, and (c) fully correlated uncertainty.
Among these, the uncorrelated uncertainty is most frequently encountered in observational data, where error bars are provided for each data point, respectively, without accounting for correlations between any pairs of data points.
We hand-write the covariance matrices in the form of multivariate Gaussians and apply them on top of the emulator that generates a stellar mass function (refer to Sec. \ref{sec:method_sim_param} for the details of the emulator). 
Subsequently, we conduct inferences with ILI using the uncertainty-equipped emulator.
In parallel, we perform Fisher analysis using the uncertainty-free emulator and the covariance matrix, respectively (refer to the last paragraph of Sec. \ref{sec:method_sim_param}). 
With this, we examine how the posterior distribution changes by imposing various correlations in the uncertainty.

\subsection{Toy Model Covariance Matrices}
We posit three toy models that have different amounts of correlation between the mass bins: i.e., zero correlation (fully uncorrelated), medium correlation (full correlation for some mass bins), and maximum correlation (fully correlated).
The covariance matrix for stellar mass functions is written as $\Sigma_{i,j} = \cov(f(M_{\star,i}),f(M_{\star,j}))$ where $f(M_{\star})$ represents the stellar mass function and $i$ represents the stellar mass bins at which it is measured, ranging from 1 to 13.
Following are the three covariance matrices with different correlations that we adopt:

\vspace{4mm}
{\bf Uncorrelated Uncertainty}\\
\begin{equation}
    \bs{\Sigma} = 
    \begin{vmatrix} 
    \Sigma_{1,1} & 0            & 0      & 0               &   0 \\
    0            & \Sigma_{2,2} & 0      & 0               &   0  \\
    0            &            0 & \ddots & 0               &   0  \\
    0            & 0            &  0     & \Sigma_{n-1,n-1}&   0  \\
    0            & 0            & 0      &  0              & \Sigma_{n,n}\\
    \end{vmatrix}.
\end{equation}
Here, $n=13$.

{\bf Partially Correlated Uncertainty}\\
The three points in the mid-mass range are fully correlated, and the rest is fully uncorrelated, namely
\begin{equation}
    \bs{\Sigma} = 
    \begin{vmatrix}
    \Sigma_{1,1} & \cdots & 0           & 0         & 0           & 0     & 0\\ 
    \vdots  & \ddots &0            & 0         & 0           & 0     & 0\\ 
    0       & 0      & \Sigma_{m-1,m-1} & \Sigma_{m,m-1} & \Sigma_{m+1,m-1} & 0     & 0\\ 
    0       & 0      & \Sigma_{m-1,m}   & \Sigma_{m,m}   & \Sigma_{m+1,m}   & 0     & 0\\ 
    0       & 0      & \Sigma_{m-1,m+1} & \Sigma_{m,m+1} & \Sigma_{m+1,m+1} & 0     & 0\\ 
    0       & 0      & 0           & 0         & 0           &\ddots & \vdots\\
    0       & 0      & 0           & 0         & 0           &\cdots & \Sigma_{n,n}
    \end{vmatrix},
\end{equation}
where $\Sigma_{i,j}=0$ for all $i, j$ where $i\neq j$ except $m-1$, $m$, and $m+1$. 
We set $m=5$, and off-diagonal components are set to $\Sigma_{i,j} = \sqrt{\Sigma_{i,i}\Sigma_{j,j}}-\epsilon$ with $\epsilon \ll 1$, imposing very high correlation among them.\\

{\bf Fully Correlated Uncertainty}\\
\begin{equation}
    \bs{\Sigma} = 
    \begin{vmatrix}
    \Sigma_{1,1} & \Sigma_{1,2} & \cdots& \Sigma_{n,1}\\ 
    \Sigma_{1,2} & \Sigma_{2,2} &\cdots & \vdots\\ 
     \vdots& \vdots & \ddots& \Sigma_{n,n-1} \\
    \Sigma_{1,n} & \cdots&\Sigma_{n-1,n} & \Sigma_{n,n}
    \end{vmatrix},
\end{equation}
where $\Sigma_{i,j} = \sqrt{\Sigma_{i,i}\Sigma_{j,j}}-\epsilon$ and $\epsilon \ll 1.$\footnote{
Note that mathematically, the fully correlated case is represented by a univariate Gaussian having a single component for the covariance matrix (i.e., standard deviation). 
However, for the consistency of the work, we use a multivariate Gaussian with a multi-dimensional covariance matrix; this, however, yields identical results to the univariate case.}
\vspace{2mm}

For simplicity, we set the components of the covariance matrices such that all diagonal elements are equal to one another (i.e., $\Sigma_{i,i}=\Sigma_{j,j}$ for all $i, j$).
Additionally, we set the diagonal elements to the same value consistently across different toy models for fair comparison.  
This guarantees that the standard deviations for the resulting stellar mass functions are identical to each other.
The standard deviation of the stellar mass functions is consistently $\sim 0.12$ dex.

\begin{figure*}[t!]
    \centering
    \includegraphics[width=\textwidth]{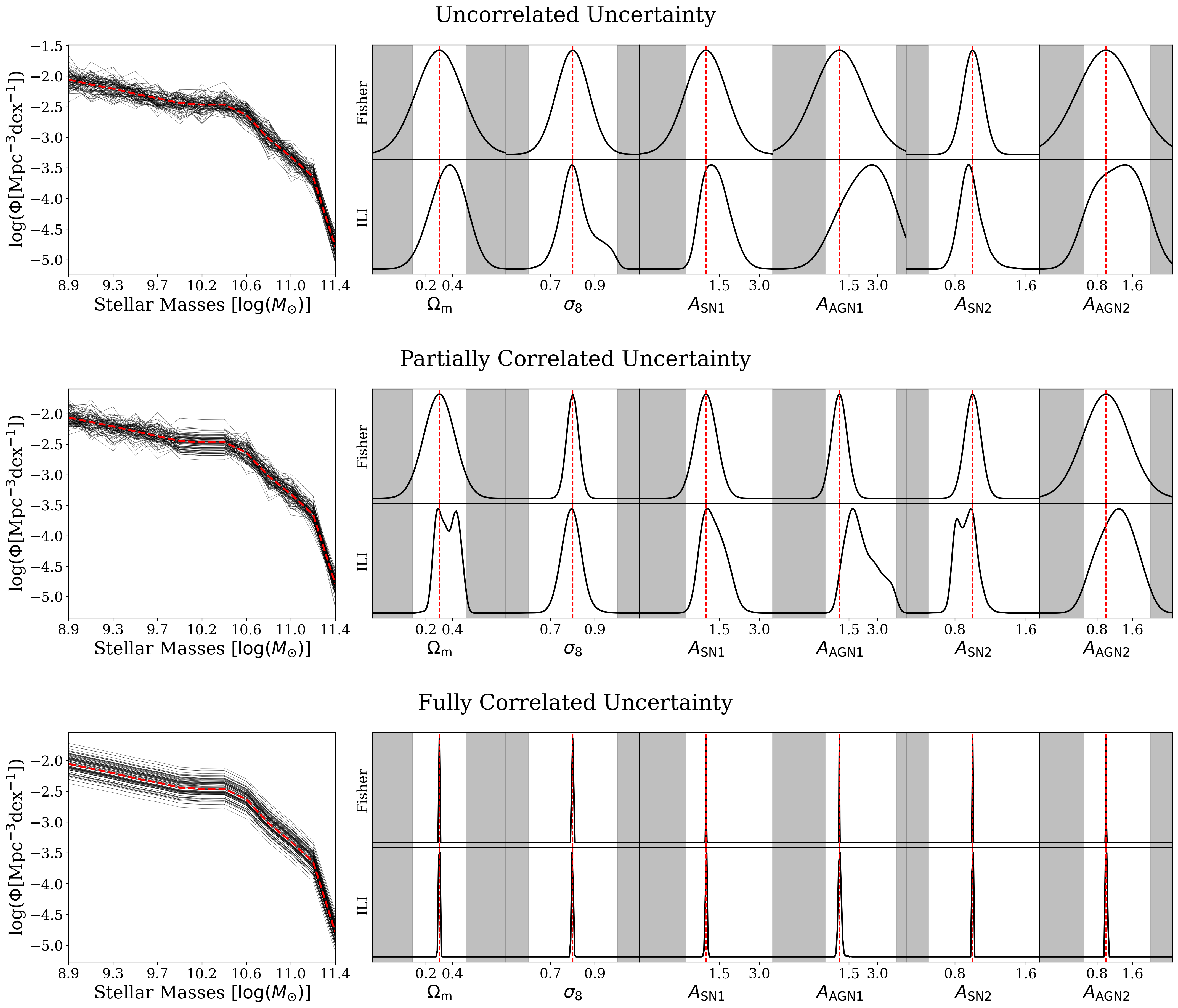}
    \caption{{\it Left}: Stellar mass functions ({\it left}) with the uncertainties from the uncorrelated uncertainty ({\it top}), the partially correlated uncertainty ({\it middle}), and fully correlated uncertainty ({\it bottom}). 
    {\it Right}: Each plot comprises the marginal distribution of parameters obtained by Fisher analysis ({\it upper}) and Implicit Likelihood Inference (ILI) ({\it lower}), respectively.
    The {\it red dashed} lines represents the target stellar mass function and parameters.
    The grey area serves as a buffer zone to facilitate visual inspection.
    }
    
    \label{fig:toy_models}
\end{figure*}

\subsection{Results and Discussion}
Fig. \ref{fig:toy_models} depicts the patterns of the uncertainties in the stellar mass functions with the aforementioned covariances ({\it left}) and the inference results ({\it right}) from Fisher analysis and ILI ({\it upper} and {\it lower} panels, respectively).
Represented in the {\it red dashed} lines are the target stellar mass function and the target parameters used to generate the target stellar mass function with the emulator.
The stellar mass functions with uncertainty (black thin curves in the left panels) are samples from the Gaussian distribution $\mathcal{N}(\bs{\mu}, \bs{\Sigma})$ with the given covariance matrix $\bs{\Sigma}$ and the mean stellar mass function from the emulator $\bs{\mu}$.
The inferences throughout this paper is performed against the target stellar mass function $\bs{\mu}_\mathrm{0}$ generated with the fiducial parameters of the CAMELS simulation suite ({\it red dashed vertical}) (refer to Sec. \ref{sec:method_sim_param} for more details).

Despite the limited prior ranges (see Sec. \ref{sec:method_sim_param}), the emulator can estimate stellar mass functions even beyond these bounds, allowing posterior distributions to extend outside the prior ranges.
However, as the emulator is trained only within the prior region, values outside this range are the results of extrapolations, which are generally considered unreliable in machine learning.
To enhance plot readability, we display results from regions outside the prior ranges as {\it shaded grey} buffer areas. 
It is worth noting that the Fisher analysis is not affected by extrapolation issues, as it provides only a variance. 
For Fisher results, we plot a Gaussian function using this variance and the fiducial parameters, whereas the ILI results represent the actual posterior distribution.

For Fisher analysis, we estimate the variance of the inferred posterior by Eq. \ref{eq:Fisher_simplified} and \ref{eq:cramer_rao}.
Presented in the {\it upper} panels of Fig. \ref{fig:toy_models} are the six univariate Gaussian distributions $\mathcal{N}(\theta_i, \sqrt{\var(\theta_{ii})})$ with the estimated variances above and the fiducial parameters as the means ({\it red dashed vertical}).
In parallel, we run the ILI framework to infer the posterior distributions of the parameters using the emulator with each covariance matrix that provides $\mathcal{N}(\bs{\mu},\bs{\Sigma})$.
Unlike the Fisher analysis, which relies on underlying assumptions such as the Gaussian distribution, ILI provides a complete posterior distribution without making such restrictive assumptions. 
This flexibility allows for a wider range of possible distributions, accounting for the noticeable differences in the peaks and shapes of the posterior distributions obtained from Fisher analysis and ILI. 
However, despite these discrepancies in the distributional characteristics, the parameter variances derived from both methods exhibit a reasonable degree of consistency with one another.

The impact of correlation on the inference is profound, as the parameter variances exhibit a diminishing trend with increasing correlation strength. 
The uncorrelated case ({\it top}), where no correlation is present, yields the broadest marginal distributions across different uncertainty realizations for almost all parameters.
The AGN parameters ($\alpha_{\rm AGN, 1}$, $\alpha_{\rm AGN, 2}$) show modest sensitivity due to its relatively weaker correlation to the stellar mass function \citep{Jo2023ApJ...944...67J}.
In the partially correlated uncertainty ({\it middle}), the overall variances become smaller compared to the uncorrelated case. 
Finally, with the introduction of full correlation ({\it bottom}), the variances collapse to distributions with negligible spread despite the same magnitude of uncertainty in the stellar mass functions.

This behavior seemingly suggests an inverse relationship between the degree of correlation present in the observables and the variance of the inferred parameters.
However, this is not necessarily the case generally, as the relationship between the degree of correlation and the inferred variance is much more complicated.
The variance of the parameters from uncertainty of observables can be determined by two main factors: 1. sensitivity of each observable with respect to the parameters, $\pd\mu_i/\pd\theta_\alpha$; 
2. correlation between data points within a model, not uncertainty, $\pd\mu_i/\pd\mu_j$.

Firstly, the variance of the parameters diminishes as the sensitivity increases. 
This phenomenon is evident in the one-dimensional case where $\var(\theta)=(\sigma\pd\theta/\pd\mu)^2$ since the sensitivity reflects how much informative the parameters are with respect to the observable.
Therefore, greater sensitivity (more informative) leads to tighter constraints (smaller variance).
Secondly, correlation between $\mu_i$ and $\mu_j$ can either increase or reduce uncertainty of the parameters.
Imagine that there exists a considerable correlation between $\mu_i$ and $\mu_j$. 
Then, regardless of the magnitude of the uncertainty associated with $\mu_j$ and $\mu_i$, the correlation itself provides additional information regarding both $\mu_i$ and $\mu_j$.
It is important to notice that the observation $\bs{\mu}_0$ is already provided, but only unknown is the certainty.
Consequently, this leads to a decrease in the variance of $\theta_\alpha$ about the uncertainty of $\mu_i$, compared to the case without correlation.
The reduction is proportional to the magnitude of $\pd\mu_j/\pd\theta_\alpha$ and correlation between $\mu_i$ and $\mu_j$.

In order to study this closely, we derive variances of parameters using Fisher analysis for the case where there are two parameters and two data points in observable (refer to Appendix \ref{apx1:derivation} for the details of the derivation).
The variance of the parameter $\theta_1$ is given as $\var(\theta_1) \geq \mathcal{F}_{11}^{-1}$ by the Cramer-Rao bound, as
\begin{equation}
\begin{split}
    \mathcal{F}_{11}^{-1} 
    &= \left(\frac{\pd\mu_1}{\pd\theta_1}\bigg|_{\theta_2}-\frac{\pd\mu_1}{\pd\mu_2}\bigg|_{\theta_1}\frac{\pd\mu_2}{\pd\theta_1}\bigg|_{\theta_2}\right)^{-2}\Sigma_{11}\\
    +& \left(\frac{\pd\mu_1}{\pd\theta_1}\bigg|_{\theta_2}-\frac{\pd\mu_1}{\pd\mu_2}\bigg|_{\theta_1}\frac{\pd\mu_2}{\pd\theta_1}\bigg|_{\theta_2} \right)^{-2} \frac{\pd\mu_1}{\pd\mu_2}\bigg|_{\theta_1} \Sigma_{12}\\
    +& \left(\frac{\pd\mu_2}{\pd\theta_1}\bigg|_{\theta_2}-\frac{\pd\mu_2}{\pd\mu_1}\bigg|_{\theta_1}\frac{\pd\mu_1}{\pd\theta_1}\bigg|_{\theta_2}\right)^{-2}\Sigma_{22}.
\end{split}
\end{equation}
The terms associated with $\Sigma_{11}$ and $\Sigma_{22}$ are relatively straightforward based on the description in the previous paragraph.
Regarding the impact of the off-diagonal elements, this result suggests that as the degree of correlation increases, the variance of the parameters not only may become smaller but also can grow larger, depending on the sign of $\pd\mu_1/\pd\mu_2|_{\theta_1}\Sigma_{12}$.
Nonetheless, it seems counter-intuitive to have a negative sign for the term, which would indicate an inverse correlation between uncertainty and the model for the observable.
Therefore, although there exists a theoretical possibility that correlation can cause impair the constraining power, it is more likely that the inclusion of covariance terms results in tighter constraints compared to the case without correlation.

The covariance terms can affect the variance of the parameters in a complex, non-linear way, leading to unique characteristics, such as bimodality or skewed distributions.
Firstly, the AGN parameters remain unaffected by the introduction of correlation when comparing the uncorrelated case to the partially correlated case. 
This can be attributed to the physical decoupling between the locus where the correlation is introduced (the mid-range of the stellar mass function) and the AGN physics, which primarily influences the high-mass end.
Also, peculiar distributional characteristics can arise due to the flexibility of ILI, and skewness is one such manifestation. 
In general, skewed distributions tend to emerge when the inferred parameter distribution exhibits a comparatively large variance relative to the domain size. 
This is because if the variance is sufficiently large, the Markov Chain Monte Carlo (MCMC) sampling process becomes unstable as it attempts to cover the entire parameter space.
Notably, the appearance of bimodality---the two peaks appear in $\om$ of partially correlated uncertainty---in response to changes in the covariance matrix is the most intriguing phenomenon observed. 
This suggests that the introduced correlation imposes constraints on the parameter space in a complex manner, originating from multiple directions simultaneously.
The bimodality will be revisited in the following section discussing the physical models.

\section{Posterior Sensitivity to the Covariance: Physically-motivated Models}
\label{sec:phys_model}

In this section, we construct various covariance matrices of uncertainties originating from physically-motivated physical models and examine their impact on posterior distributions. 
We utilize two different observational data. 
The first consists of accumulated posteriors from 491 3D-HST galaxies \citep{skelton2014ApJS..214...24S} fit with {\tt Prospector-$\alpha$} \citep{leja2019ApJ...877..140L}. 
With this dataset, we approximate the effects of sampling on the covariance matrix. 
The second observation is inferred from a single observed galaxy by varying parameters for the stellar initial mass function. 
As the stellar initial mass function is nearly unobservable for the great majority of galaxies \citep{conroy2012ApJ...760...71C} but can easily have up to a +/-0.5 dex effect on the inferred mass \citep{wang2024ApJ...963...74W}, perhaps the simplest assumption is that all galaxies have the same IMF but that IMF is unknown. We use this single object as a way to construct this model.
Lastly, we investigate the effect of the observational aperture size using the IllustrisTNG100-1 simulation. 
This involves varying the surface brightness threshold that determines the radius within which the stellar mass of galaxies is calculated.

\begin{figure*}[t!]
    \centering
    \includegraphics[width=\textwidth]{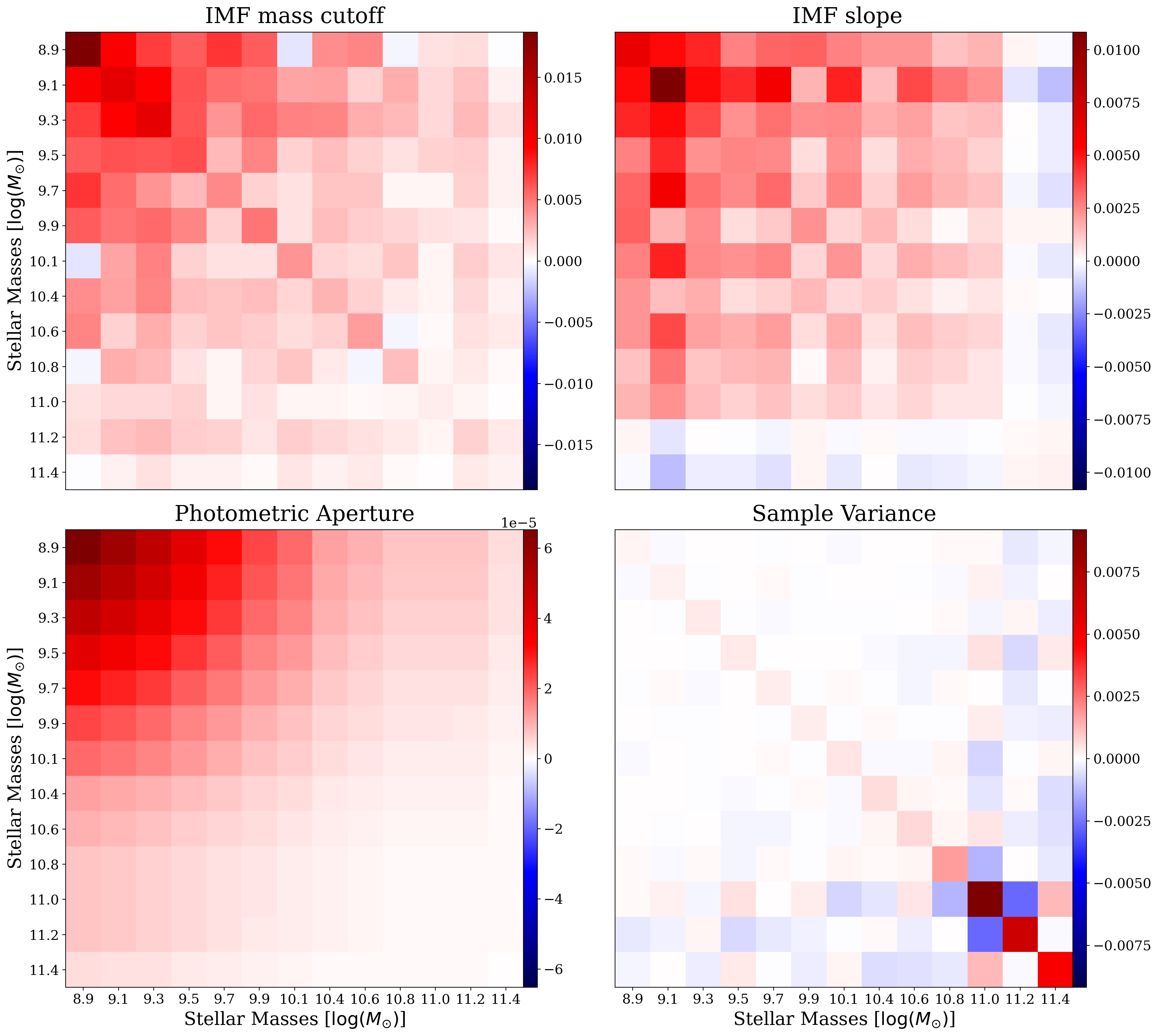}
    \caption{Heat maps of the covariance matrices.
    The positive elements of the covariances are colored in `blue', while the negative elements are colored in `red'.}
    \label{fig:cov_heatmap}
\end{figure*}

\subsection{Construction of Covariance Matrices}
\label{sec:phys_model_matrices}
The physics-driven covariance in this work represents the correlation within observation-driven physical quantities (e.g., galaxy stellar mass function) with respect to uncertainty of (physical) parameters in physical models (e.g., SED fitting), through which observation data is processed to obtain the observation-driven physical quantity.
Given the observation data $f_\nu$ (such as observed photometry) of a single galaxy, the probability distribution of stellar mass, marginalized over a set of various contributing factors, $\bs{h}$, (such as IMF) can be rigorously written as:
\begin{equation}
\label{eq:smf}
    p(M_{\star}|{f_{\nu}}) = \int{p(M_{\star}|f_{\nu},\bs{h})p(\bs{h})\dd\bs{h}}.
\end{equation}
This is the probability distribution of the stellar mass of a single galaxy given the observation data of the galaxy.
Then, the galaxy stellar mass function can be obtained by integrating this over the observational curve space $f_{\nu}$ with proper normalization (in other words, integration over observed galaxies):
\begin{equation}
\label{eq:smf}
    p(M_{\star}) = C\int{p(M_{\star}|f_{\nu},\bs{h})p(\bs{h})p(f_{\nu})\dd f_{\nu}\dd\bs{h}},
\end{equation}
where $C$ is the normalization factor.
This represents the stellar mass function incorporating uncertainty marginalizing over every possible source.

However, we can also introduce the galaxy stellar mass function conditioned on contributing parameters $\bs{h}$ as
\begin{equation}
    p(M_{\star}|\bs{h}) = C\int{p(M_{\star}|f_{\nu},\bs{h})p(f_{\nu})\dd f_{\nu}},
\end{equation}
which represents how the galaxy stellar mass functions change in response to the parameters.
This eventually yields the correlations in the galaxy stellar mass functions with respect to the parameters.
Subsequently, in order to capture the correlations arising from each parameter, we isolate one parameter $h_t$ from a set of parameters $\bs{h}$ by re-writing Eq. \eqref{eq:smf} as
\begin{equation}
    p(M_{\star}|h_t) = \int{p(M_{\star}|h_t, f_{\nu}, \bs{h}')p(f_\nu)p(\bs{h}')\dd{f_\nu}\dd\bs{h}'}.
\end{equation}
where $\bs{h}' \cup \{h_t\} = \bs{h}$ and $\bs{h}' \cap \{h_t\} =\O$. 
Here, a bold letter stands for a `set' of parameters, whereas a regular letter represents an individual `parameter'.
With this, we construct the covariance matrix $\tilde{C}_{ij}$ of uncertainty with respect to $\bs{h}_t$ between two arbitrary stellar mass bins $M_i$ and $M_j$:
\begin{align}
\label{eq:covariance}
     \tilde{C}_{ij} =&\, \cov_{h_t}(M_i,M_j)\\
     =&\, \mathbb{E}_{h_t}\big\langle\left[p(M_i|h_t)-\mathbb{E}_{h_t}\big\langle p(M_i|h_t)\big\rangle\right]\\
     &\times\left[p(M_j|h_t)-\mathbb{E}_{h_t}\big\langle p(M_j|h_t)\big\rangle\right]\big\rangle,
\end{align}
where $\mathbb{E}_{h_t}\langle p(M_i|h_t)\rangle = \int_{h_t\in \mathcal{H}_t} p(M_i|h_t)p(h_t)\dd h_t$. 
Here, $\mathcal{H}_t$ is the space where $h_t$ lives, and the integral should cover the entire space.

{\bf Slope \& high-mass cutoff for IMF}\\
It is typical to assume a fixed IMF when fitting galaxy spectral energy distributions, as described in Sec. \ref{sec:method_observation}. 
That being said, constructing a covariance matrix from a single galaxy requires several oversimplified assumptions.

The first assumption posits that the posterior distribution's configuration from an observation curve is independent of the galaxy stellar mass, maintaining a constant pattern. 
This enables us to replicate posterior distributions across various stellar mass values.
That is, the probability distribution of stellar mass conditioned on an arbitrary $f_{\nu}$ in terms of $f_{\nu,0}$ can be written as
\begin{equation}
\begin{split}
p(M_{\star}|&\alpha, f_{\nu}, \bs{h}')=\\
&p\Big(M_{\star}-\left(\bar{M}(f_\nu)-\bar{M}_0\right)\Big\vert\alpha, f_{\nu,0}, \bs{h}'\Big)
\end{split}
\end{equation}
where $\bar{M}(f_{\nu})$ is the mean galaxy stellar mass and $\alpha$ stands for the slope of IMF.
Here, $\bar{M}_0=\bar{M}(f_{\nu,0})$ is of the actual observed galaxy.
Then, the stellar mass function can be written as 
\begin{equation}
\label{eq:smf_2}
\begin{split}
p(M_{\star}|\alpha) &=\\
\int&p\Big(M_{\star}-\left(\bar{M}(f_\nu)-\bar{M}_0\right)\Big\vert\alpha, f_{\nu,0}, \bs{h}'\Big)\\
&\times p(f_\nu)\dd f_{\nu}\dd \bs{h}'.
\end{split}
\end{equation}

The second assumption is that the number of galaxies in each stellar mass bin in the SMF can be approximated using the galaxy population of the CAMELS simulations with the fiducial parameters.
In Eq. \ref{eq:smf_2}, the integration over $f_\nu$ effectively represents summing all observed galaxies, each weighted by $f(\nu)$. 
However, this is impractical as we only have data for a single galaxy. 
This assumption allows us to replace the integration over the observational data of a single galaxy ($f_\nu$) with a summation over $M'$, where $M'$ represents bin masses sampled from the CAMELS simulations.
Thus, with $p(f_\nu)df_\nu \approx N(M')/V$,
\begin{equation}
\begin{split}
\tilde{p}&(M_{\star}|\alpha) = \frac{1}{V}\sum_{M'}\Big[ N(M')\\
&\times \int 
p\Big(M_{\star}-\left(M'-\bar{M}_0\right)\Big\vert\alpha, f_{\nu,0}, \bs{h}'\Big)\dd \bs{h}'\Big],
\end{split}
\end{equation}
where  $N(M')$ and $V$ are the number of galaxies in the $M'$ bins and the volume of the CAMELS simulations, respectively.
Note that $\tilde{p}$ is not a probability distribution, but a physical entity.
By plugging $\tilde{p}(M_\star|\alpha)$ into Eq. \ref{eq:covariance}, we obtain the covariance matrix \footnote{The covariance matrix we obtained from the data is not positive definite due to numerical accuracy. 
To resolve this issue, we added the identity matrix multiplied by a negligibly small number.}.
We repeat the same procedure for the high-mass cut parameter and derive its covariance matrix.
\\

{\bf Mock Photometric Aperture}\\
To mimic observational uncertainty related to photometric aperture threshold within the simulation environment, we utilize the IllustrisTNG100-1 simulation. 
We achieve this by varying the surface brightness cutoff used to estimate the stellar masses of galaxies and computing the corresponding stellar mass function (SMF) in the TNG100-1 simulation.
Surface density of galaxy at a certain radius is computed as follows:
We project the star particles onto two dimension;
We run through all the star particles of each galaxy and find N nearest star particles with respect to their radial distance from the center of mass of the galaxy;
With this, we can forms a ring with a certain width depending on the average distance amongst the star particles within the ring;
Then, we estimate its surface density by calculating the flux of the rings by summing up the fluxes of each particle with the g band of stellar photometry that IllustrisTNG provides \citep[for details]{nelson2018MNRAS.475..624N}.
We use 100 different thresholds for surface density cutoffs: $\mu_\mathrm{thres}=[-10,-20]$ with a linear spacing where $\mu_\mathrm{thres}=-2.5\log_{10}(f_{g, \mathrm{ring}})$ with $f_{g, \mathrm{ring}}$ being the flux of a ring.
We determine the size of galaxies to be the outermost radius of at which the surface density of a galaxy meets the surface density thresholds.
Above all, we obtained 100 different stellar mass estimates for each galaxy.
Subsequently, we calculate the covariance matrix, which approximately accounts for the uncertainty arising from aperture size.
We have the detailed discussion in Appendix \ref{apx:aperture}.
\\

{\bf Sample Variance}\\
We construct the covariance accounting for the sample variance by utilizing the Cosmic Variance\footnote{In cosmology, cosmic variance is unambiguously characterized as ``the inevitable contribution to the variance of parameter estimates arising from the fact that we are confined to observations of a single Universe.'' 
Concurrently, the CV set represents the variance derived from the observation (or sampling) of a limited volume of the universe. 
Henceforth, we will refer to it as `sample variance'.} (CV) set of the CAMELS simulations \citep{camels2021ApJ...915...71V}.  
The CV set consists of 27 simulations with a co-moving volume of $(25\, \mathrm{Mpc}/h)^3$ with fixed cosmology and astrophysics that sample `sample variance' using different initial conditions;
The estimation of covariance is conducted in a straightforward manner.
We generate the 27 galaxy stellar mass functions from each simulation in the CV set. 
With 27 stellar mass functions denoted as $f_n(M_i)$ where n stands for different stellar mass functions, ranging from 1 to 27, and $M_i$ is a mass bin, we calculate the covariance as $\tilde{C}_{ij}=\mathbb{E}\big\langle[ f_n(M_i)-\mathbb{E}\langle f_n(M_i)\rangle][ f_n(M_j)-\mathbb{E}\langle f_n(M_j)\rangle]\big\rangle$ where $\mathbb{E}\langle f_n \rangle = \sum_{n=1}^{27}f_n/27$.
Subsequently, we re-scale the covariance matrix to match the uncertainty size of the galaxy stellar mass functions to that of simulations of $(75 \mathrm{Mpc}/h)^3$ as in Fig. 2 (d) of \citet{genel2014MNRAS.445..175G}.

\begin{figure*}[t!]
    \centering
    \includegraphics[width=\textwidth]{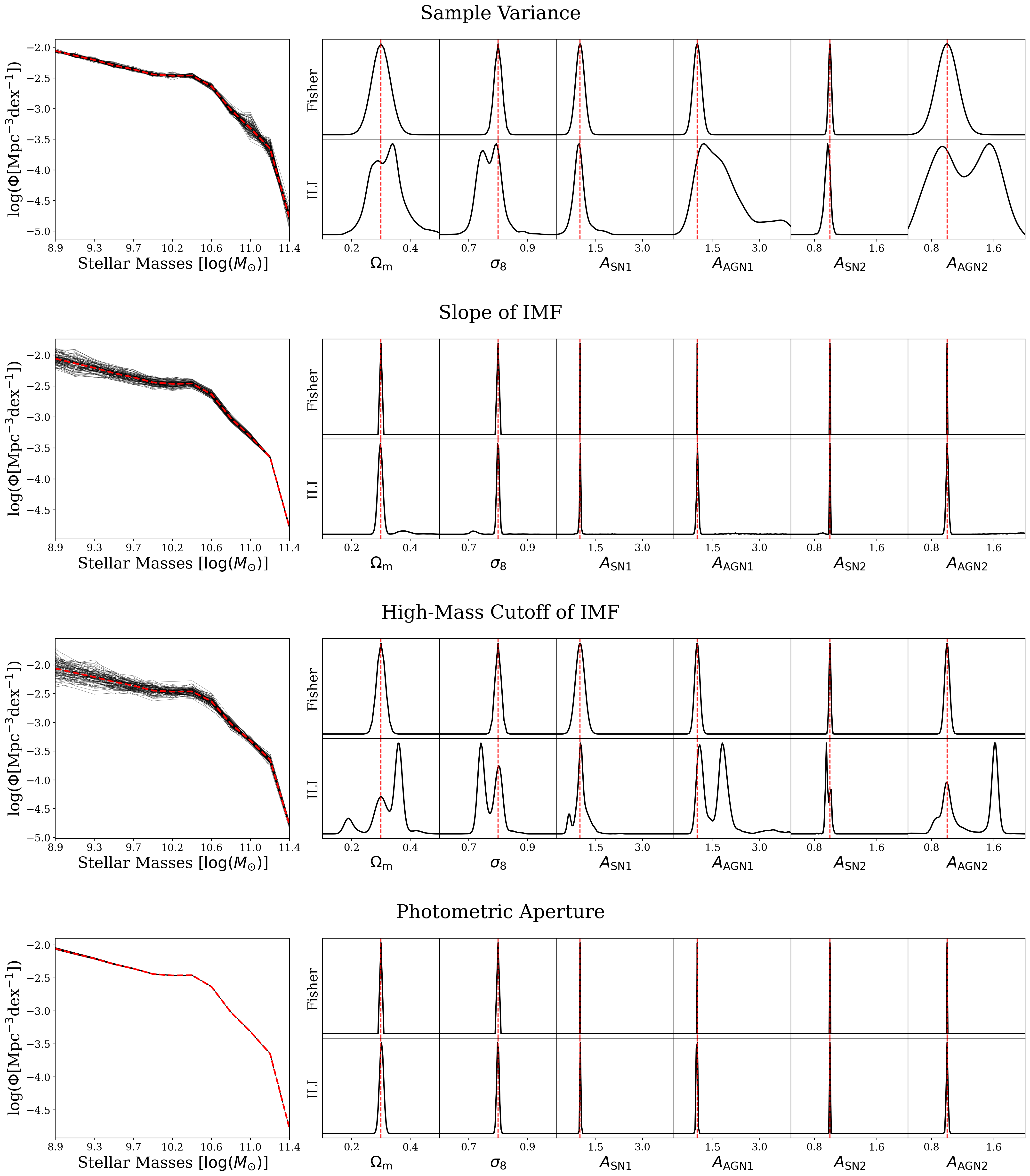}
    \caption{{\it Left}: Stellar mass functions ({\it left}) with the uncertainties from the uncorrelated uncertainty ({\it top}), the partially correlated uncertainty ({\it middle}), and fully correlated uncertainty ({\it bottom}). 
    {\it Right}: Each plot comprises the marginal distribution of parameters obtained by Fisher analysis ({\it upper}) and Implicit Likelihood Inference (ILI) ({\it lower}), respectively.
    The {\it red dashed} lines represents the target stellar mass function and parameters.}
    \label{fig:phys_model}
\end{figure*}

\begin{figure*}
    \includegraphics[width=\textwidth]{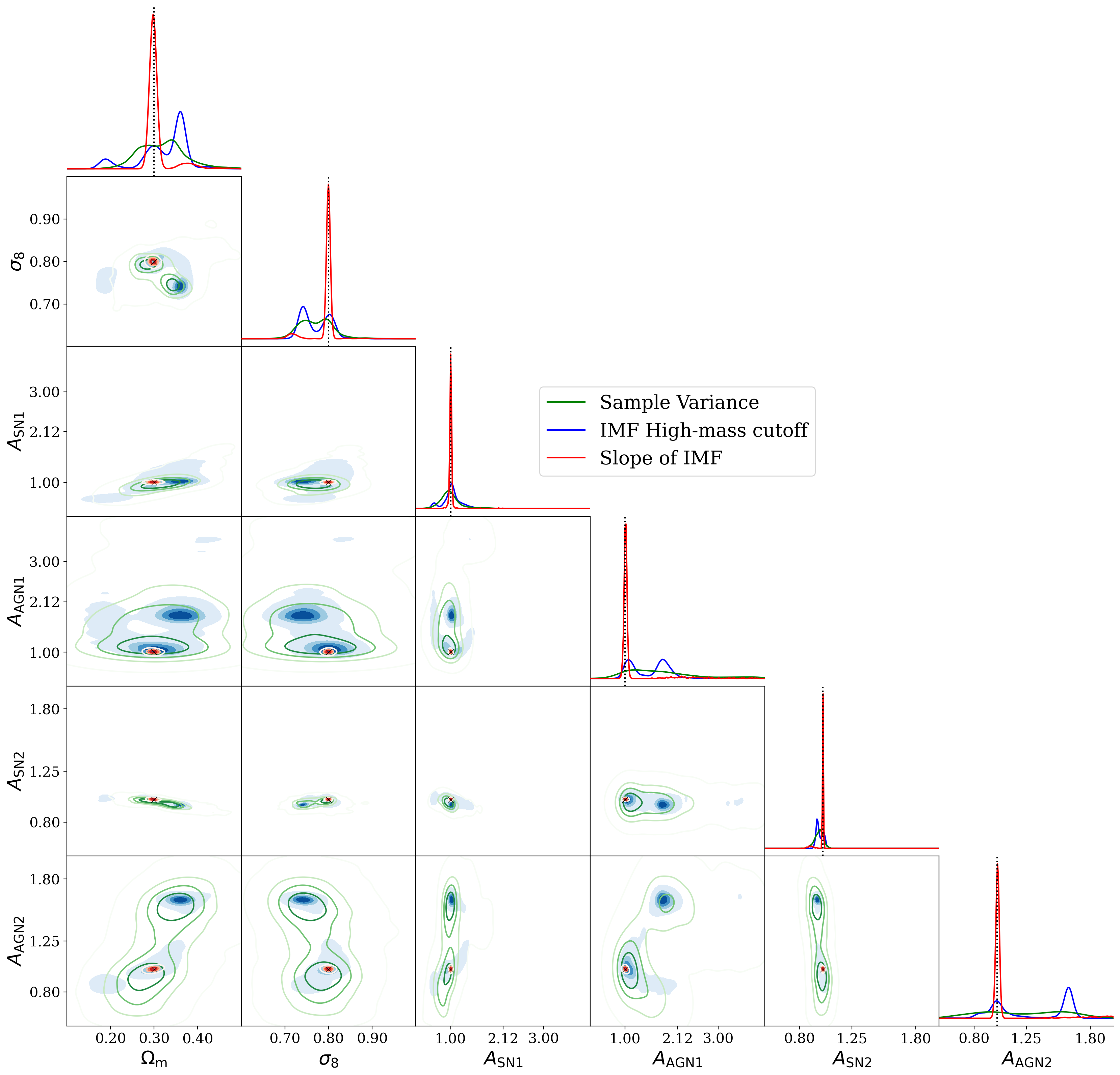}
    \caption{Two dimensional projection ({\it lower left}) and marginals ({\it diagonal}) of the posterior distributions of the sample variance ({\it green}), the high-mass cutoff of the IMF ({\it blue}), and the slope of the IMF ({\it red}).
    The {\it black} lines and crosses represent the true parameters.
    }
    \label{fig:projection}
\end{figure*}

\subsection{Result and Discussion}
\label{sec:phys_model_results}

Fig. \ref{fig:cov_heatmap} illustrates the visualization of the covariance matrices that were computed in the previous section and used in the inferences.
The color maps are customized for each case such that `red' and `blue' represent positive and negative values---or correlations, respectively, except for photometric aperture, which displays positive values only.
Each individual pixel represents one of the 13 distinct mass bins.
Overall, the variances of lower mass end are affected more than higher. 
The photometric aperture exhibits a consistent trend where the covariance diminishes towards the higher mass ends.
Furthermore, each off-diagonal element seemingly shows similarity to $\Sigma_{i,j} = \sqrt{\Sigma_{i,j}\Sigma_{j,j}}$, which is analogous to the behavior seen in the fully correlated case.
The covariances of the IMF mass cut and IMF slope show a comparable pattern of higher correlations at the lower mass end, which diminish towards the higher mass end. 
However, the IMF mass cut distinctly displays some anti-correlations colored in light blue, whereas IMF the slope is characterized by anti-correlations between higher mass bins and both the lower and middle mass bins.
In contrast, the sample variance presents relatively strong diagonal elements with the coherent, yet small off-diagonal elements at lower mass end, akin to the uncorrelated case in the toy models.
Notably, the higher mass end exhibits a significant anti-correlation and stronger diagonal elements, stemming from the high variance at the high mass end due to lack of massive galaxies in the simulations. 

Fig. \ref{fig:phys_model} shows the patterns of the uncertainties in the stellar mass functions with the covariances from Sec.  \ref{sec:phys_model_matrices} ({\it left}) and the inference results ({\it right}) from Fisher analysis and ILI ({\it upper} and {\it lower} panels, respectively), analogously to the structure of Fig.~\ref{fig:toy_models}.
{\it Red dashed} lines are the target stellar mass function and the target parameters used to generate the target stellar mass function with the emulator.
A notable difference from Sec. \ref{sec:toy_models} is that the (non-zero) values in the covariance matrices and the corresponding uncertainties on the stellar mass function values are not set to the same constant value across different models but are rather derived directly from the physically-motivated models.

The variations in covariance matrices introduce intricate and non-linear constraints on the posterior distributions, exhibiting a more complex behavior compared to Sec. \ref{sec:toy_models}. 
The sample variance ({\it first row}) shows relatively larger uncertainty of the stellar mass function, particularly at the high mass end. 
This results in considerable degeneracy and variances notably in $\aagnone$, and $\aagntwo$, which are often considered responsible for population of massive galaxies e.g., via AGN feedback. 
Meanwhile, the $\asnone$ and $\asntwo$ are relatively better constrained, suggesting that relatively small uncertainty at the lower mass end could be influential.
In the case of the slope of IMF ({\it second row}), it exhibits a relatively simple result, having narrow one peaks for each parameter.
In contrast, the high-mass cutoff of the IMF ({\it third row}) exhibits highly complex posterior distributions compared to other cases. 
The high-mass cutoff of the IMF inferred via ILI does not simply yield broader posteriors compared to the slope of the IMF but rather exhibits complicated multi-modal distributions, whereas the Fisher analysis results (by construction) in monotonous single-peaked distributions.

Moreover, while the sample variance exhibits uncertainty that is comparable to or marginally less than that of the IMFs, it demonstrates substantially greater variance in the parameters compared to the IMFs.

The comparison between the slope of the IMF and the high-mass cutoff of the IMF reveals an interesting yet relatively straightforward pattern.
Despite the comparable uncertainty in the stellar mass functions, the variances of the posterior distributions in the former case are notably smaller than in the latter. 
Furthermore, while the sample variance exhibits uncertainty that is comparable to or marginally less than that of the IMFs, it demonstrates substantially greater variance in the parameters compared to the IMFs.
This aligns with the findings presented in Sec. \ref{sec:toy_models}, which demonstrate how the variances of the parameters change in response to correlations in the uncertainty.
Notably, these results highlight that, in some cases, the impact of correlation on inference can outweigh that of the magnitude of uncertainty itself.
In a different light, incorporating information about covariance can provide an additional dimension for parameter constraints, underscoring the importance of accounting for covariance in observation.
We revisit and further explore these posterior distributions below through two-dimensional projection plots.

Lastly, the covariance associated with the aperture effect ({\it fourth row}) is highly similar to the fully correlated case, which typically yields negligible parameter variances. 
A detailed examination of the aperture effect, given as an approximately fully correlated case, is provided in Appendix \ref{apx:aperture}.
The relatively small variances observed for the IMF slope and aperture effect, compared to the fully correlated case shown in Fig. \ref{fig:toy_models}, suggest that some physical parameters may empirically lead to covariance matrices with correlations that are nearly as strong as the theoretical fully correlated case, which leads to negligible impacts on inference.

Fig. \ref{fig:projection} shows two-dimensional projections of the posterior distributions for the sample variance ({\it Green}), the high-mass cutoff ({\it blue}), and the slope of the IMF ({\it red}), with the true parameter values marked as {\it black} lines and crosses. 
The posterior distribution of the sample variance covers a a large portion of parameter space, exhibiting a lower probability in contrast to other significant peaks.
For the high-mass cutoff case, it is evident that there are two prominent high-density regions and one smaller peak in the $\om$ dimension.
One of the mass cut peaks coincides with the peaks of the the slope of the IMF and the true parameter values. 
However, despite sharing the same locus, the posterior distribution for the the slope of the IMF exhibits a distinct orientation and elongation, differing from the mass cut case.
This observation highlights that the characteristics of the covariance matrix can have an indispensable impact on the statistical properties of the inferred posterior distributions, such as their shape, orientation, and the presence of multiple modes or peaks.

\section{Summary}
\label{sec:sum}
We investigate the impact of correlations between observational uncertainties, namely the covariance between observables, on model parameter inference in the context of galaxy population observables and cosmological simulations. 
Our study focuses on stellar mass functions as the observable, with varying degrees of correlation between the mass bins, and six cosmological and astrophysical parameters from the CAMELS simulations as the parameters of interest (details in Sec. \ref{sec:method_sim_param}).

We employ an emulator trained on CAMELS simulations as a forward model to generate stellar mass functions from input parameters. 
We then perform inference using two distinct approaches. 
The first is Fisher analysis, a classical method for measuring variances, which involves constructing a Fisher information matrix (Eq. \ref{eq:Fisher_simplified}) and estimating variance using the Cramér-Rao bound (Eq. \ref{eq:cramer_rao}). 
The second approach is Implicit Likelihood Inference (ILI), which utilizes machine learning to train a neural network that estimates posterior distributions in a Bayesian framework (refer to Sec. \ref{sec:method_ili}).
By applying both Fisher analysis and ILI, we aim to ensure consistency in our results and compare the outcomes of these different inference approaches. 

To account for correlations in uncertainty, we construct covariance matrices and use them to generate Gaussian noise, which in turn produces stellar mass functions with uncertainty.
First, we construct three simple and conceptually straightforward covariance matrices that can be easily written by hand: fully correlated, partially correlated, and uncorrelated (Sec. \ref{sec:toy_models}). Setting the diagonals for these matrices identically, but the non-diagonal elements (i.e.~covariances) differently, allows us to understand how different degrees of correlation in uncertainty affect inference.
Next, we use actual observational data to create more realistic covariance matrices that capture the uncertainties associated with various astrophysical and observational factors (Sec. \ref{sec:phys_model}). 
These covariance matrices incorporate correlations arising from IMF parameters in the SED fitting, aperture size, and sample variance. 
This helps us investigate how intricate and practical covariances impact the inference results.

In scenarios driven by physical motivations, although the uncertainties associated with the stellar mass functions due to sample variance are smaller compared to those related to the slope of the IMF and are comparable, the variances of the posterior distributions pertaining to the sample variance are substantially greater than those associated with the slope of the IMF.

The toy models are designed to investigate the impact of covariance in a straightforward way.
As a result, we find that the correlation within the uncertainty tends to reduce the inferred variances of the parameters, occasionally leading to distinctive features such as bimodality or skewed distributions. 
However, we also demonstrate that theoretical possibility that the inclusion of correlation in uncertainty can indeed increase the variances.
In the physics-motivated cases, although the uncertainties of with the stellar mass functions from the sample variance are smaller compared to those from the slope of the IMF, the variances of the posterior distributions regarding the sample variance are substantially greater than those associated with the slope of the IMF.
Moreover, the slope of the IMF resolves the degeneracy in the high-mass cutoff of the IMF, despite the comparable magnitude of uncertainties in the stellar mass functions.
This underscores that, in certain cases, the impact of correlation on inference can surpass that of the magnitude of uncertainty.
The inference with high-mass cutoff of the IMF reveals intricate multi-modal distributions, with one of the peaks aligning with the peaks derived from the slope of the IMF and the truth (target parameters). 
This suggests the potential that incorporating correlation can resolve the issue of degeneracy.
Overall, the inclusion of covariance terms in the uncertainty can have profound impact on the inference.


Further notable observations are made:
\begin{itemize}
    \item  The amount of correlation present in the observables is generally inversely proportional to the variance of the inferred parameters. Higher correlation leads to tighter constraints and narrower marginal distributions. 

    \item Simple Fisher analysis overall agrees with ILI in terms of the variances, lending support to the ILI results.
    That being said, ILI's non-parametric nature provides flexibility in morphology of posterior distributions, allowing for capturing intricate features like multi-modality and skewness, which may be missed by Fisher analysis assumptions. 
    
    \item The mass cut of IMF exhibited highly complex, multi-modal posterior distributions, despite having variances (namely, diagonal elements of the covariance matrix) of approximately the same magnitude as those arising from the slope of the IMF, which resulted in simpler, broader posteriors.
    
    \item The characteristics of the covariance matrix, such as the presence of correlation and its structure, can have an indispensable impact on the statistical properties of the inferred posterior distributions, including their shape, orientation, and the presence of multiple modes or peaks.
    
    \item  Two-dimensional projections of the posterior distributions revealed distinct features, such as multiple high-density regions and elongations, highlighting the intricate and non-linear effects of different covariance matrices on the posterior distributions.
    
\end{itemize}

One caveat should be addressed:
Although the results regarding the dynamic relationship between covariance and posterior remains valid, the physical results and quantitative insights presented in this work are fundamentally contingent upon the fidelity of both the underlying physical models---cosmological simulations, emulators, and observational data processing techniques. 
Further work is required to rigorously characterize the physical relationship between covariance and inference results.

This work emphasizes the significant role played by the covariance matrix---specifically in uncertainty of galaxy stellar mass functions---in shaping the posterior distributions and the importance of accounting for its characteristics in characterizing uncertainties and interpreting parameter estimates, utilizing cosmological simulations and its parameters.
Consequently, precise covariance modeling offers additional constraining capabilities for parameter inference, potentially enabling the resolution of degeneracy within posterior distributions.
In the field of the galaxy population observation, it has traditionally been uncommon to incorporate covariance terms in observable.
However, we anticipate that the implementation of a rigorous covariance matrix can yield novel physical insights.

\acknowledgments
The Flatiron Institute is supported by the Simons Foundation.

\appendix

\section{Impact of correlation terms in the Cramer-Rao bound}
\label{apx1}
\subsection{Derivation of Variance}
\label{apx1:derivation}
In order to mathematically estimate the impact of correlation within uncertainty, we analytically derive variance bounds using the Cramer-Rao bound and Fisher information.
For simplicity, we focus on the case of two observables (say, two bins in the mass function) $\bmu=(\mu_1, \mu_2)$ and two model parameters $\btheta=(\theta_1, \theta_2)$.
Imagine that we have a general form of the positive definite square covariance matrix
\begin{equation}
    \Sigma =  \begin{bmatrix}
        \Sigma_{11} & \Sigma_{12}\\
        \Sigma_{12} & \Sigma_{22}
    \end{bmatrix},
\end{equation}
where $\det(\Sigma) = \Sigma_{11}\Sigma_{22}-\Sigma_{12}^2>0$ and $\Sigma = \Sigma^{\top}$.

Starting from the Fisher matrix $\mathcal{F}_{ij} = \frac{\pd\bmu^{\top}}{\pd\theta_i} \Sigma^{-1} \frac{\pd\bmu}{\pd\theta_j}$, and using a notation where $\pd_i\mu_\alpha = \frac{\pd\mu_\alpha}{\pd\theta_i}$,
\begin{equation}
\begin{split}
    \mathcal{F}_{ij} &= \frac{\pd\bmu^{\top}}{\pd\theta_i} \left(\frac{1}{\det(\Sigma)}\begin{bmatrix}
        \Sigma_{22} & - \Sigma_{12}\\
        -\Sigma_{12} & \Sigma_{11}
    \end{bmatrix}\right) \frac{\pd\bmu}{\pd\theta_j}\\
    &= 
    \frac{1}{\det(\Sigma)}
    \begin{bmatrix}
        \pd_i\mu_1 & \pd_i\mu_2
    \end{bmatrix}
    \begin{bmatrix}
        \Sigma_{22} & - \Sigma_{12}\\
        -\Sigma_{12} & \Sigma_{11}
    \end{bmatrix}
    \begin{bmatrix}
        \pd_j\mu_1\\
        \pd_j\mu_2
    \end{bmatrix}\\
    &= \frac{1}{\Sigma_{11}\Sigma_{22}-\Sigma_{12}^2}\left(
     \Sigma_{11}\pd_i\mu_2\pd_j\mu_2
    -\Sigma_{12} (\pd_i\mu_1\pd_j\mu_2+\pd_i\mu_2\pd_j\mu_1) 
    +\Sigma_{22}\pd_i\mu_1 \pd_j\mu_1
    \right).
\label{eq:Fij}
\end{split}
\end{equation}
Then,
\begin{equation}
\begin{split}
    \mathcal{F}_{ij}^{-1} &= \frac{1}{\det(\mathcal{F})}\mathrm{adj} (\mathcal{F}_{ij})\\
    &=\left(\frac{1}{n!}\epsilon_{i_1i_2\cdots i_n}\epsilon_{j_1j_2\cdot j_n} \mathcal{F}_{i_1j_1} \mathcal{F}_{i_2j_2}\cdots \mathcal{F}_{i_nj_n}\right)^{-1}
    \left(
    \frac{1}{(n-1)!}\epsilon_{ii_2\cdot i_n}\epsilon_{jj_2\cdot j_n}\mathcal{F}_{i_2j_2}\cdots \mathcal{F}_{i_nj_n}
    \right).
\end{split} 
\end{equation}
Here, the Levi-Civita symbol $\epsilon$ and the Einstein convention are adopted, where $\epsilon_{i_1,i_2,\cdots,i_n}$ is $+1$ for an even permutation, $-1$ for an odd permutation, and $0$ for any two indices are identical.
For the two-by-two symmetric matrix, this can be simplified as
\begin{equation}
\begin{split}
    \mathcal{F}_{ij}^{-1} &= \frac{1}{\mathcal{F}_{11}\mathcal{F}_{22}-\mathcal{F}_{12}^2}\epsilon_{i\alpha}\epsilon_{j\beta}\mathcal{F}_{{\alpha}{\beta}}.
\end{split}
\label{eq:Finv2x2}
\end{equation}
To calculate $\det(\mathcal{F})$, we use Equation \ref{eq:Fij} to expand the first and second terms in the denominator in Equation \ref{eq:Finv2x2} as
\begin{equation}
\begin{split}
\mathcal{F}_{11}\mathcal{F}_{22} &= 
\frac{1}{(\Sigma_{11}\Sigma_{22}-\Sigma_{12}^2)^2}
\left(\Sigma_{11}(\pd_1\mu_2)^2-2\Sigma_{12}(\pd_1\mu_1\pd_1\mu_2)+\Sigma_{22}(\pd_1\mu_1)^2\right) 
\left(\Sigma_{11}(\pd_2\mu_2)^2-2\Sigma_{12}(\pd_2\mu_1\pd_2\mu_2)+\Sigma_{22}(\pd_2\mu_1)^2\right)\\
\end{split} 
\end{equation}
and 
\begin{equation}
\begin{split}
\mathcal{F}_{12}^2 &= 
\frac{1}{(\Sigma_{11}\Sigma_{22}-\Sigma_{12}^2)^2}
\left(\Sigma_{11}(\pd_1\mu_2\pd_2\mu_2)-\Sigma_{12}(\pd_1\mu_1\pd_2\mu_2+\pd_1\mu_2\pd_2\mu_1)+\Sigma_{22}(\pd_1\mu_1\pd_2\mu_1)\right)^2,\\
\end{split} 
\end{equation}
respectively. Then, the denominator in Equation \ref{eq:Finv2x2} is given by
\begin{equation}
\begin{split}
\mathcal{F}_{11}\mathcal{F}_{22} -\mathcal{F}_{12}^2 &= 
\frac{1}{(\Sigma_{11}\Sigma_{22}-\Sigma_{12}^2)^2}
\Big[\\
&
\Sigma_{11}^2((\pd_1\mu_2\pd_2\mu_2)^2-(\pd_1\mu_2\pd_2\mu_2)^2)\\
+&
\Sigma_{22}^2((\pd_1\mu_1\pd_2\mu_1)^2-(\pd_1\mu_1\pd_2\mu_1)^2)\\
+&
\Sigma_{11}\Sigma_{22}\left((\pd_1\mu_2\pd_2\mu_1)^2+(\pd_1\mu_1\pd_2\mu_2)^2-2(\pd_1\mu_1\pd_2\mu_2\pd_1\mu_2\pd_2\mu_1)\right)\\
+&
2\Sigma_{11}\Sigma_{12}\left(-((\pd_1\mu_2)^2\pd_2\mu_1\pd_2\mu_2+(\pd_2\mu_2)^2(\pd_1\mu_1\pd_1\mu_2))+(\pd_1\mu_2\pd_2\mu_2(\pd_1\mu_1\pd_2\mu_2+\pd_1\mu_2\pd_2\mu_1))\right)\\
+&
2\Sigma_{12}\Sigma_{22}\left(-\pd_1\mu_1\pd_1\mu_2(\pd_2\mu_1)^2-(\pd_1\mu_1)^2\pd_2\mu_1\pd_2\mu_2+(\pd_1\mu_1\pd_2\mu_1)(\pd_1\mu_1\pd_2\mu_2+\pd_1\mu_2\pd_2\mu_1)\right)\\
+&
\Sigma_{12}^2(4(\pd_1\mu_1\pd_2\mu_2\pd_1\mu_2\pd_2\mu_1)-(\pd_1\mu_1\pd_2\mu_2+\pd_1\mu_2\pd_2\mu_1)^2)
\Big]\\
&=
\frac{1}{(\Sigma_{11}\Sigma_{22}-\Sigma_{12}^2)^2}
\left[(\Sigma_{11}\Sigma_{22}-\Sigma_{12}^2)(\pd_1\mu_2\pd_2\mu_1-\pd_1\mu_1\pd_2\mu_2)^2\right]\\
&=
\frac{1}{\Sigma_{11}\Sigma_{22}-\Sigma_{12}^2}
\left[(\pd_1\mu_2\pd_2\mu_1-\pd_1\mu_1\pd_2\mu_2)^2\right]
\end{split} 
\end{equation}
Plugging this into each element of $\mathcal{F}^{-1}$, which is the lower bound of the variance of each parameter according to the Cramer-Rao bound (see Eq. \ref{eq:cramer_rao}),
\begin{equation}
\label{eq:apx_f11}
\begin{split}
    \mathcal{F}_{11}^{-1} &= 
    \frac{1}{\mathcal{F}_{11}\mathcal{F}_{22}-\mathcal{F}_{12}^2}
    \mathcal{F}_{22}\\
    &=
    \frac{1}{(\pd_1\mu_2\pd_2\mu_1-\pd_1\mu_1\pd_2\mu_2)^2}
    \left(\Sigma_{11}(\pd_2\mu_2)^2-2\Sigma_{12}(\pd_2\mu_1\pd_2\mu_2)+\Sigma_{22}(\pd_2\mu_1)^2\right),
\end{split}
\end{equation}
\begin{equation}
\begin{split}
    \mathcal{F}_{22}^{-1} &= 
    \frac{1}{\mathcal{F}_{11}\mathcal{F}_{22}-\mathcal{F}_{12}^2}
    \mathcal{F}_{11}\\
    &=
    \frac{1}{(\pd_1\mu_2\pd_2\mu_1-\pd_1\mu_1\pd_2\mu_2)^2}
    \left(\Sigma_{11}(\pd_1\mu_2)^2-2\Sigma_{12}(\pd_1\mu_1\pd_1\mu_2)+\Sigma_{22}(\pd_1\mu_1)^2\right),
\end{split}
\end{equation}
\begin{equation}
\begin{split}
    \mathcal{F}_{12}^{-1} &= 
    -\frac{1}{\mathcal{F}_{11}\mathcal{F}_{22}-\mathcal{F}_{12}^2}
    \mathcal{F}_{12}\\
    &=
    -\frac{1}{(\pd_1\mu_2\pd_2\mu_1-\pd_1\mu_1\pd_2\mu_2)^2}
    \left(\Sigma_{11}(\pd_1\mu_2\pd_2\mu_2)-\Sigma_{12}(\pd_1\mu_1\pd_2\mu_2+\pd_1\mu_2\pd_2\mu_1)+\Sigma_{22}(\pd_1\mu_1\pd_2\mu_1)\right).
\end{split}
\end{equation}
Thus, we can conclude that correlation in the uncertainties does not necessarily lead to greater uncertainty in the target parameters.
For instance, when $2\Sigma_{12}(\pd_1\mu_1\pd_1\mu_2) < 0$, the inferred variance $\mathcal{F}^{-1}_{11}$ can be greater than in the uncorrelated case ($\Sigma_{12}=0$).
This outcome depends not only on the correlations in uncertainty $\Sigma_{12}$ but also on the model's sensitivity to the parameters $\pd\mu/\pd\theta$.
Although this derivation is specific to the two-dimensional case, it can be generalized to the N-dimensional scenario.

\subsection{Implication of correlation in the variance of parameters}
\label{apx1:interpretation}
We have derived a complicated expression for the variances of each parameter.
In this section, we focus on the physical interpretation of $\mathcal{F}_{11}^{-1}$ and how the covariance terms work in the inference.

Firstly, we revert to the one-dimensional scenario in which a single observable and a single parameter exist (i.e., $\mathcal{F} =(\pd\mu/\pd\theta)^2\sigma^{-2}$).
In this case, the variance is given by $\var(\theta)=(\sigma\pd\theta/\pd\mu)^2$ where $\sigma$ is the standard deviation of observable.
This is relatively easy to interpret. 
$\sigma$ is the standard deviation of the observable $\mu$, and $\pd\mu/\pd\theta$ represents how much the observable should change in response to changes in $\theta$, which links $\theta$ to $\mu$, and vice versa.
Thus, $\sigma\times\pd\theta/\pd\mu$ translates the uncertainty of observable $\sigma$ into the uncertainty of $\theta$ through the susceptibility of $\theta$ to $\mu$, that is, $\pd\theta/\pd\mu$.

We decompose $\mathcal{F}_{11}^{-1}$ from Eq. \ref{eq:apx_f11} to three parts $a_{11}$, $a_{12}$, $a_{22}$ associating with $\Sigma_{11}$, $\Sigma_{12}$, and $\Sigma_{22}$ such that $\mathcal{F}_{11}^{-1} = a_{11} + a_{12} + a_{22}$.
Then, the first term $a_{11}$ associated with $\Sigma_{11}$ is written as
\begin{equation}
\begin{split}
    a_{11} &= \frac{(\pd_2\mu_2)^2}{(\pd_1\mu_2\pd_2\mu_1-\pd_1\mu_1\pd_2\mu_2)^2}\Sigma_{11} \\
    &= \left(\pd_1\mu_1-\frac{\pd_1\mu_2\pd_2\mu_1}{\pd_2\mu_2}\right)^{-2}\Sigma_{11} \\
    &=  \left(\frac{\pd\mu_1}{\pd\theta_1}\bigg|_{\theta_2}-\frac{\pd\mu_1}{\pd\theta_2}\bigg|_{\theta_1}\frac{\pd\mu_2}{\pd\theta_1}\bigg|_{\theta_2}\frac{\pd\theta_2}{\pd\mu_2}\bigg|_{\theta_1}\right)^{-2}\Sigma_{11}\\
    &=  \left(\frac{\pd\mu_1}{\pd\theta_1}\bigg|_{\theta_2}-\frac{\pd\mu_1}{\pd\mu_2}\bigg|_{\theta_1}\frac{\pd\mu_2}{\pd\theta_1}\bigg|_{\theta_2}\right)^{-2}\Sigma_{11}.\\
\end{split}
\end{equation}
$a_{11}$ represents how $\theta_1$ should vary in response to $\Sigma_{11}$, which represents the uncertainty of $X_1$---the first element of the observable.
The first term with in the parenthesis is straightforward, which is the same as the one-dimensional case---$\var(\theta)=(\sigma\pd\theta/\pd\mu)^2$.

The second term yields the contribution from correlation of mean of $\bs{X}={\mu_1, \mu_2}$.
To provide insight, consider two coins---$Y_1$ and $Y_2$---linked by a special string such that when flipping them, they are highly likely to have the same side.
Also assume that we can only ascertain their sides with a certain probability (or uncertainty) of $\bs{p}={p_1, p_2}$ due to the blur vision.
Consequently, when flipping them, regardless of uncertainty of $p_1$, the outcome of $Y_2$ can affect the determination of the $Y_1$ outcome due to the correlation between $Y_1$ and $Y_1$.
Returning to our analysis, this effect can be represented by the term $\pd\mu_1/\pd\mu_2$. 
Subsequently, this term affects the unceratinty of $\theta_1$ via $\pd\mu_2/\pd\theta_1$. 
As a side note, $a_{22}$ can be derived by simply interchanging the subscripts 1 and 2 of $\mu$.

For $a_{12}$ associated with $\Sigma_{12}$, we can write
\begin{equation}
\begin{split}
    a_{12} &= \frac{\pd_2\mu_1\pd_2\mu_2}{(\pd_1\mu_2\pd_2\mu_1-\pd_1\mu_1\pd_2\mu_2)^2}\Sigma_{12}\\
    &=\left(\frac{(\pd_1\mu_2\pd_2\mu_1)^2-2\pd_1\mu_2\pd_2\mu_1\pd_1\mu_1\pd_2\mu_2+(\pd_1\mu_1\pd_2\mu_2)^2}{\pd_2\mu_1\pd_2\mu_2}\right)^{-1}\Sigma_{12}\\
    &=\left(\frac{(\pd_1\mu_2)^2\pd_2\mu_1}{\pd_2\mu_2}-2\pd_1\mu_2\pd_1\mu_1+\frac{(\pd_1\mu_1)^2\pd_2\mu_2}{\pd_2\mu_1}\right)^{-1}\Sigma_{12}\\
    &=\left[\frac{\pd\mu_2}{\pd\theta_1}\bigg|_{\theta_2}^2\frac{\pd\mu_1}{\pd\theta_2}\bigg|_{\theta_1}\frac{\pd\theta_2}{\pd\mu_2}\bigg|_{\theta_1}-2\frac{\pd\mu_2}{\pd\theta_1}\bigg|_{\theta_2}\frac{\pd\mu_1}{\pd\theta_1}\bigg|_{\theta_2}+\frac{\pd\mu_1}{\pd\theta_1}\bigg|_{\theta_2}^2\frac{\pd\mu_2}{\pd\theta_2}\bigg|_{\theta_1}\frac{\pd\theta_2}{\pd\mu_1}\bigg|_{\theta_1} \right]^{-1}\Sigma_{12}\\
    &=\left[(\frac{\pd\mu_2}{\pd\theta_1}\bigg|_{\theta_2}^2\frac{\pd\mu_1}{\pd\mu_2}\bigg|_{\theta_1}-2\frac{\pd\mu_2}{\pd\theta_1}\bigg|_{\theta_2}\frac{\pd\mu_1}{\pd\theta_1}\bigg|_{\theta_2}+\frac{\pd\mu_1}{\pd\theta_1}\bigg|_{\theta_2}^2\frac{\pd\mu_2}{\pd\mu_1}\bigg|_{\theta_1} \right]^{-1}\Sigma_{12}\\
    &= \left(\frac{\pd\mu_1}{\pd\theta_1}\bigg|_{\theta_2}-\frac{\pd\mu_1}{\pd\mu_2}\bigg|_{\theta_1}\frac{\pd\mu_2}{\pd\theta_1}\bigg|_{\theta_2} \right)^{-2} \frac{\pd\mu_1}{\pd\mu_2}\bigg|_{\theta_1} \Sigma_{12}.
\end{split}
\end{equation}
We can notice that $a_{12}$ is almost identical to $a_{11}$ except for the factor $\pd\mu_1/\pd\mu_2|_{\theta_1}$.
This factor along with $\Sigma_{12}$ reflects correlation in uncertainty between $X_1$ and $X_2$.
The contribution of the covariance term to $\mathcal{F}_{11}^{-1}$ is determined by the sign of $\frac{\pd\mu_1}{\pd\mu_2}\big|_{\theta_1}\Sigma_{12}$.
However, when we have a positive correlation ($\Sigma_{12}>0$), $\mu_1$ and $\mu_2$ should change in the opposite direction (i.e., $\frac{\pd\mu_1}{\pd\mu_2}\big|_{\theta_1}>0$), and vice versa, by definition.
In other words, $a_{12}$ should be negative at all times.
Consequently, the presence of correlation in the uncertainty unequivocally enhances the precision of inference by diminishing the variance.

\begin{figure}[H]
    \centering
    \includegraphics[width=0.50\linewidth]{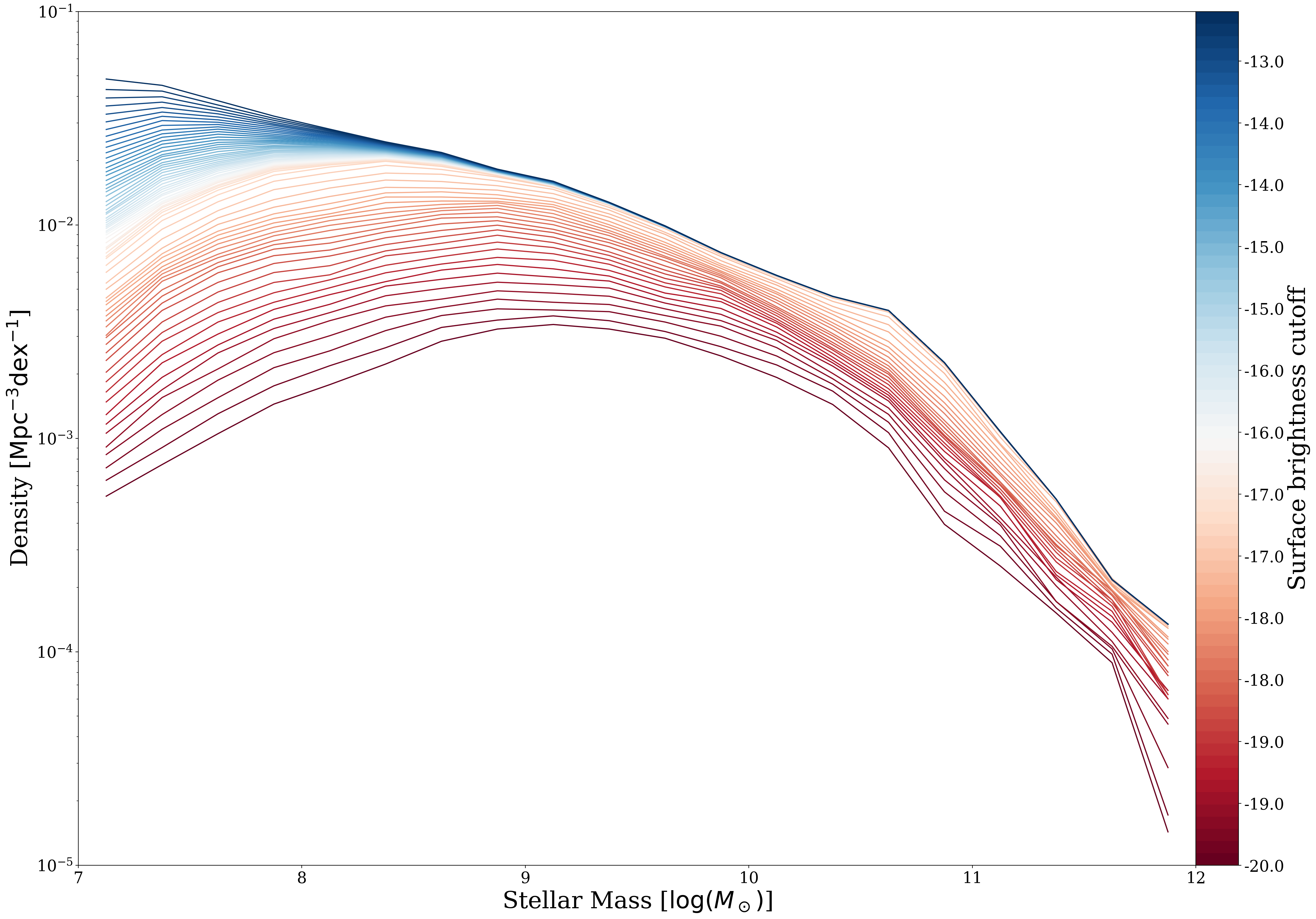}
    \caption{Stellar mass functions with respect to the surface brightness cutoffs.
    A higher surface brightness cutoff is represented by the blue color, whereas a lower surface brightness cutoff is illustrated by the red color. }
    \label{fig:SMF_surface_brightness}
\end{figure}

\begin{figure}[htbp]
    \centering
    \includegraphics[width=0.50\linewidth]{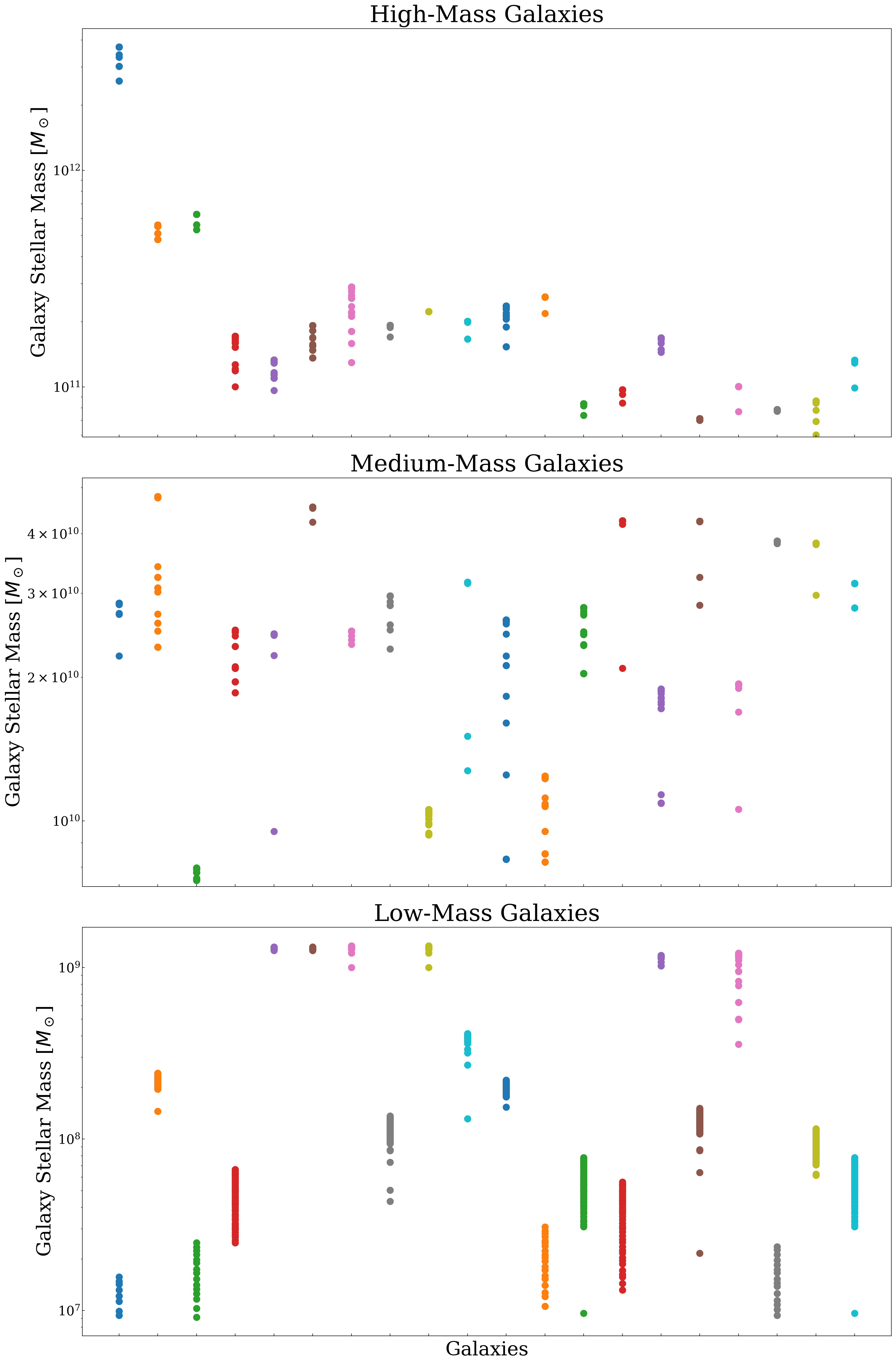}
    \caption{Decline of the galaxy stellar masses with respect to the surface brightness cutoffs for 20 galaxies for high, middle, and low mass ranges, respectively.}
    \label{fig:stellar_mass_surface_brightness}
\end{figure}

\section{Galaxy Stellar Mass with Photometric Aperture}
\label{apx:aperture}
In this section, we present the details of how the galaxy stellar mass of IllustrisTNG100-1 changes with respect to the surface brightness thresholds.
To begin, we compute the surface brightness of galaxies of the IllustrisTNG100-1 as follows:
1. We project the star particles onto two dimensions;
2. For each galaxy, we find 10 nearest star particles with respect to their radial distance from the center of mass of the galaxy, resulting in rings with a certain width depending on the average distance amongst the star particles within the ring;
3. We estimate the surface density of each ring by calculating the flux of the rings by summing up the fluxes of each particle with $g$-band stellar photometry that IllustrisTNG provides \citep[for details]{nelson2018MNRAS.475..624N}.
The flux of a ring $f_{g, \mathrm{ring}}$ is calculated as
\begin{equation}
\begin{split}
f_{g, \mathrm{ring}} &= \sum_{n}  f_{g, n}\\
&= \sum_{n} 10^{-0.4*m_g},
\end{split}
\end{equation}
where $f_{g, n}$ and $m_g$ are a flux of each star particle and the magnitude of the g band;
4. We use 100 different thresholds for surface density cutoffs: $\mu_\mathrm{thres}=[-10,-20]$ with a linear spacing where $\mu_\mathrm{thres}=-2.5\log_{10}(f_{g, \mathrm{ring}})$ with $f_{g, \mathrm{ring}}$ being the flux of a ring.
We determine the size of galaxies to be the outermost radius of at which the surface density of a galaxy meets the surface density thresholds.
Above all, we obtained 100 different stellar mass estimates for each galaxy.
Subsequently, we calculate the covariance matrix, which approximately accounts for the uncertainty arising from aperture size.

Fig. \ref{fig:SMF_surface_brightness} depicts the galaxy stellar mass functions with respect to the surface brightness cutoff.
A higher surface brightness cutoff is represented by the blue color, whereas a lower surface brightness cutoff is illustrated by the red color.
In the red regime, while lower mass end show a constant decrease, the mid-high mass range show dense, stacked region, which results from high mass galaxies moving horizontally to the left as surface brightness cutoff increases.
In the blue regime, although the lower mass end shows a consistent decline as the surface brightness cutoff increases, the higher mass end remains largely unaffected until a surface brightness cutoff of $\sim -16$. 
This phenomenon can be attributed to the substantial mass of high-mass galaxies, which makes them less susceptible to mass loss as the surface brightness cutoff increases.

Within the red regime, the constant decrease is shown at the lower mass end, while the mid-high mass end displays a more moderate decline.
This can be attributed to the distinctive surface density profile of the galaxies.
Fig. \ref{fig:stellar_mass_surface_brightness} illustrates the reduction in stellar masses of galaxies corresponding to increasing surface brightness cutoffs.
This reveals that the magnitude of decrease in mass is unique for each galaxy, depending on its surface brightness profile.
In addition, the low-mass galaxies tend to show the steep decreases compared to low and high mass galaxies.
Lastly, as depicted in Fig. \ref{fig:SMF_surface_brightness}, the densely stacked lines at the higher mass end suggest that an increase in surface brightness cutoff results in a horizontal shift to the left for high mass galaxies, losing their masses, which is in line with intuition.

\bibliography{main}{}
\bibliographystyle{aasjournal}

\end{document}